\documentclass{aa}

\usepackage{graphicx}

\begin{document}

\title{On the efficiency of particle acceleration by rotating magnetospheres in AGN}

\author
{ Zaza Osmanov \inst{1} \fnmsep\thanks{On leave from Evgeni
Kharadze Georgian National Astrophysical Observatory, Kazbegi
ave.~2a, Tbilisi-0160, Georgia} \and Andria Rogava \inst{2,3}
\fnmsep\thanks{On leave from Evgeni Kharadze Georgian National
Astrophysical Observatory, Kazbegi ave.~2a, Tbilisi-0160, Georgia}
\and Gianluigi Bodo \inst{4} }

\offprints{Z. Osmanov}

\institute { Dipartimento di Fisica Generale, Universit\'a degli
Studi di Torino, Via Pietro Giuria 1, Torino I-10125, Italy\\
\email{osmanov@to.infn.it} \and Centre for Plasma Astrophysics,
K.U.Leuven, Celestijnenlaan 200B, 3001 Leuven, Belgium\\
\email{Andria.Rogava@wis.kuleuven.be} \and Abdus Salam
International Centre for Theoretical Physics, Trieste
I-34014,Italy\\ \email{arogava@ictp.it} \and Osservatorio
Astronomico di Torino, Strada dell'Osservatorio 20, I-10025, Pino
Torinese, Italy
\\
\email{bodo@to.astro.it} }

\date{}

\abstract
{}
{To investigate the efficiency of centrifugal acceleration of
particles as a possible mechanism for the generation of ultra-high
$\gamma$-ray nonthermal emission from TeV blazars, we study the
centrifugal acceleration of electrons by rotating magnetic field
lines, for an extended range of inclination angles and determine
the maximum Lorentz factors $\gamma_{max}$ attainable by the
electrons via this process.}
{Two principal limiting mechanisms for the particle acceleration,
inverse Compton scattering and breakdown of the bead-on-the-wire
approximation, are examined.}
{Particles may be centrifugally accelerated up to $\gamma_{max}
\simeq 10^8$ and the main limiting mechanism for the
$\gamma_{max}$ is the inverse Compton scattering.}
{The energy of centrifugally accelerated particles can be amply
sufficient for the generation (via inverse Compton scattering) of
the ultra-high energy (up to $20TeV$) gamma emission in TeV
blazars.}

\keywords{ Galaxies: active,jets,nuclei; Radiation mechanisms:
non-thermal; Gamma rays: theory; Galaxy: center. }

\maketitle
%
%________________________________________________________________

\section{Introduction}

One puzzling and interesting problem, related to active galactic
nuclei (AGNs) is the origin of the ultra-high $\gamma$-ray
emission from blazars. Some blazars, such as Mrk 421, Mrk 501, PKS
2155-304, 1ES 2344-514, H1426+428, and 1ES 1959+650, emit TeV
($1TeV \equiv 10^{12}eV$) photons and form a special class of
``TeV blazars" (\cite{d02,cw99,h03}).

The standard blazar model implies the presence of a supermassive
black hole, surrounded by an accretion disk and ejecting twin
relativistic jets, one of which is seen almost end-on. The            %'end-on' or 'face-on', which is correct?%
broadband emission spectrum of blazars is made of two components:
the low-energy (from radio to optical/UV) part attributed to
synchrotron radiation and the high-energy (from X-rays to
$\gamma$-rays) component formed by the inverse Compton scattering
(ICS) of softer photons (\cite{ktk02,ketal98}). The latter part of
the spectrum is usually interpreted on the basis of the
synchrotron self-Compton (SSC) model. However, the origin of
accelerated and/or pre-accelerated electrons and the mechanism
that is responsible for their efficient acceleration is still a
matter of uncertainty. Proposed mechanisms such as the Fermi-type
acceleration process (\cite{cw99}) and re-acceleration of
electron-positron pairs as a feedback mechanism (\cite{ghb93}) may
account for the observed high energy emission up to 20 TeV.
However, the Fermi-type acceleration in relativistic jets is
efficient when the seed population of "pre-accelerated" electrons
possesses quite high ($\gamma_{max}\ge10^2$) Lorentz factors
(\cite{rm00}). The origin of this "pre-acceleration" is still a
matter of discussion.

Centrifugally driven outflows (CDOs), centrifugally accelerated
particles may acquire quite high energies. This concept
(\cite{g68,g69}) has often been discussed
(\cite{mr94,chetal96,g96,ckf99}) in the context of pulsar emission
theory. Regarding the AGNs it was shown that CDOs from accretion
disks occurred if the poloidal magnetic field lines are inclined
at an angle $\le 60^0$ to the equatorial plane of the disk
(\cite{bp82}).

In the AGN context, the presence of the CDO would imply that
despite the intense UV radiation (when via ICS, soft photons are
scattered against accelerated electrons and, as a result,
electrons lose energy and photons gain energy), electrons may
reach quite high, $\gamma_{max} \sim 10^5$, Lorentz factors. If
efficient enough, this mechanism could be used not only to justify
the pre-acceleration of electrons but it could be considered as an
alternative mechanism to generate Blazar high-energy emission (up
to $\sim 20TeV$) (\cite{oku}).

Using the approach suggested by Gangadhara (\cite{g96}), where a
test charge motion along a rapidly rotating field line was
considered in the context of a millisecond pulsar, Gangadhara and
Lesch examined the role of the centrifugal force on the dynamics
of electrons moving along straight magnetic field lines, fixed in
the equatorial plane and co-rotating with the spinning AGN
(\cite{gl97}). They have shown that scattering of low-energy
photons against accelerated electrons may lead to the generation
of the nonthermal X-ray and $\gamma$-ray emission.

This problem was re-examined by Rieger and Mannheim (\cite{rm00})
who tried to specify whether the rotational energy gain of charged
particles, moving along straight magnetic field lines, is limited
not only by ICS but also by the breakdown of the-bead-on-the-wire
approximation (BBW). The latter would happen in the vicinity of
the light cylinder when a Coriolis force acting on the particle
and trying to `tear it off' the field line  would exceed the
Lorentz force binding the particle to the field line. According to
their consideration, the maximum value of the Lorentz factor,
being the subject of both limitations, is $\gamma_{max}\sim1000$,
which is not enough to produce the ultra-high energy photons
emitted by TeV blazars.

In real, three-dimensional astrophysical situations (for example
jets) the magnetic field lines are not localized in the equatorial
plane but are inclined with respect to it. In this paper we
re-examine the same problem considering the wide range of possible
inclinations. We show that, for a wide range of AGN, the mechanism
responsible for limiting the attainable maximum Lorentz factors is
ICS, which under certain conditions can allow particles to reach
quite high Lorentz factors $\gamma_{max} \ge 10^{5}$. The BBW
becomes important for the low luminosity ($<10^{41}erg/s$) AGN,
when $\gamma_{max}\sim10^8$. For higher luminosities
($>10^{41}erg/s$) it can be dominant only for relatively small
inclinations of the magnetic field lines with respect to the
rotation axis ($\leq10^\circ$). Therefore, contrary to the
conclusions of Rieger and Mannheim (\cite{rm00}), we argue that
CDOs could be efficient enough to account for the TeV blazar
emission.

The paper is arranged in the following way: In Sec. 2 we derive
basic equations. In Sec. 3 we consider a case when magnetic field
lines are located in the equatorial plane and show that the
presence of the ICS allows a wide range of AGN to generate
ultra-high photon emission. In Sec. 4 inclined magnetic field
lines are considered. We show that in this case ICS still remains
the dominant limiting mechanism for most of the AGN. In the final
section we summarize the obtained results.

%__________________________________________________________________

\section{Main consideration}

We consider a typical AGN with a central black hole mass
$M_{BH}=10^8M_{\bigodot}$ and an angular rate of rotation $\omega
\sim 3\times 10^{-5}s^{-1}$ (that is only 1\% of the maximum
possible rotation rate for black holes with this mass), which
makes the light cylinder radius located at $r_L\approx 10^{15}
cm$. The values of the angular velocity and corresponding light
cylinder radius are typical for AGN winds (\cite{belv,rm00}). This
spinning rate corresponds to the Keplerian radius $\sim8\times
R_g$ (where $R_g \equiv 2GM_{BH}/c^2$ is the gravitational radius)
which is about where softer photons are expected to originate
(\cite{peterson}); these photons are instrumental for the Inverse
Compton mechanism.

Rieger \& Mannheim (2000) consider two limiting mechanisms for the
maximum Lorentz factor: ICS, when soft photons are scattered
against electrons gaining energy and suspending further electron
acceleration and BBW when due to the violation of the balance
between the Coriolis and Lorentz forces, acting on the electron,
it leaves the magnetic field line.

When the electron moves along the rotating magnetic field line it
experiences the centrifugal force and as a result accelerates. The
corresponding time scale describing the acceleration process can
be defined as (\cite{rm00}):

$$ t_{acc}\equiv \frac{\gamma}{d\gamma/dt}. \eqno(1) $$

The acceleration lasts until the electron encounters a photon,
which may limit the maximum Lorentz factor of the particle. This
process may be characterized by the corresponding cooling time
scale  (\cite{Lightman},\cite{rm00}):

$$ t_{cool}=3\times 10^7\frac{\gamma}{(\gamma^2-1)U_{rad}}[s],
\eqno(2) $$ where $U_{rad}=\tau l_e\times L_{Edd}/4\pi c R^2$ is
the energy density of the radiation, $\gamma$-the Lorentz factor
of the electron, $\tau\leq1$, $L_{Edd}$ is the Eddington
luminosity (for the mentioned mass of AGN $L_{Edd}\approx
10^{46}erg/s$), $l_e \equiv L/L_{Edd}$ ($L$ is a disk luminosity),
$10^{-7} \leq l_e\leq1$. In Rieger \& Mannheim (2000) the range
$10^{-4}\leq l_e\leq1$ was used but in this paper we consider the
whole range of AGN luminosities (down to AGN with the lowest
luminosities $10^{39}erg/s$ ($l_e \approx 10^{-7}$) (\cite{AGN}).

Let us start by the general case when straight magnetic field
lines are inclined  by the angle $\theta$ to the rotation axis and
supposing that magnetic field lines remain straight up to the
light cylinder.  Let us introduce a metric in the co-moving frame
of reference $$ ds^2=-c^2\left(1-\frac{\omega^2
R^2sin^2\theta}{c^2}\right)dt^2+dR^2,\eqno(3a) $$ derived from the
Minkowskian metric ($ds^2={\eta}_{\alpha
\beta}dx^{\alpha}dx^{\beta}$, with ${\eta}_{\alpha \beta} \equiv
diag\{-1,+1,+1,+1 \}$ and $x^{\alpha}\equiv (ct;x,y,z)$) after the
following variable transformation: $x=Rsin\theta cos\omega t$,
$y=Rsin\theta sin\omega t$ and $z=Rcos\theta$.

Defining $\Omega=\omega sin\theta$ (3a) reduces  to: $$
ds^2=-c^2\left(1-\frac{\Omega^2
R^2}{c^2}\right)dt^2+dR^2,\eqno(3b) $$ which formally coincides
with the metric in the co-moving frame of reference in the case of
the inclination angle $\theta=90^\circ$ (\cite{mr94}). This means
that all kinematic expressions valid for the equatorial plane are
also valid for the inclined  trajectories of electrons if instead
of $\omega$ we use $\Omega$. For the equation of motion, from
(3b), we get: $$
 \frac{d}{d\tau}\frac{\partial
L}{\partial\dot{\bar{x}}^{\alpha}}=\frac{\partial
L}{\partial\bar{x}^{\alpha}},\eqno(4a) $$ $$
L=-\frac{1}{2}mc\bar{g}_{\alpha
\beta}\frac{d\bar{x}^{\alpha}}{d\lambda}\frac{d\bar{x}^{\beta}}{d\lambda}\eqno(4b)
$$ $$ \bar{g}_{\alpha \beta} \equiv
diag\left\{-\left(1-\frac{\Omega^2
R^2}{c^2}\right),1\right\},\eqno(4c) $$ $$ \bar{x}^{\alpha}\equiv
(ct;R),
\dot{\bar{x}}^{\alpha}\equiv\frac{d\bar{x}^{\alpha}}{d\lambda}.\eqno(4d)
$$

According to this approach all physical quantities are functions
of a parameter $\lambda$. Then from Eqs. (4a) for $\alpha = 0$
using the four velocity identity
$\bar{g}_{\alpha\beta}(d\bar{x}^{\alpha}/d\lambda)(d\bar{x}^{\beta}/d\lambda)=-1$
one can express the Lorentz factor of electrons as a function of
the radial distance (\cite{mr94}): $$ \gamma =  \frac{1}{
\sqrt{\widetilde{m}} \left(1-\frac{
\Omega^2R^2}{c^2}\right)}.\eqno(5) $$

Combining Eqs. (1) and (5), for the acceleration time scale we
have (\cite{rm00}): $$ t_{acc} =
\frac{c\sqrt{1-\frac{\Omega^2R^2}{c^2}}}{2\Omega^2R\sqrt{1-\widetilde{m}
\left(1-\frac{ \Omega^2R^2}{c^2}\right)} },\eqno(6) $$ where
$\widetilde{m} =(1-\Omega^2R_0^2/c^2-v_0^2/c^2)/(
1-\Omega^2R_0^2/c^2)^2$. $R_0$ and $v_0$ are initial position and
initial radial velocity of the particle, respectively.

To evaluate the efficiency of each of the limiting mechanisms one
needs the expression for the maximum Lorentz factor attainable by
an electron subject to  ICS. Initially the electrons accelerate
and this process lasts until the energy gain is balanced by the
energy losses due to ICS. This happens when $t_{acc} \simeq
t_{cool}$. From Eqs. (2), (5), and (6) it follows that near the
light cylinder, when $\Omega R/c \rightarrow 1$ (and $\gamma
\rightarrow \infty$) both time scales tend to zero, which means
that ICS starts working efficiently in the vicinity of the light
cylinder, hampering the subsequent acceleration of the electron.
Using the corresponding condition for time scales: $$ t_{cool}
\approx t_{acc} \eqno(7) $$ and combining Eqs. (2), (6), and (7)
one may easily derive an approximate expression for the maximum
Lorentz factor: $$ \gamma^{ICS}_{max}\approx
10^{14}\sqrt{\widetilde{m}}\left[\frac{ 6\Omega }{U_{rad}(R_L)
}\right]^2, \eqno(8) $$ where $R_L\approx r_L/sin\theta$.

In order to derive the analogous estimation of the maximum Lorentz
factor limited by the BBW mechanisms we can use the method
developed by Gangadhara (\cite{g96}). For this purpose we use the
force responsible for the BBW. The particle momentum in the
laboratory frame of reference is: $$ {\bf P} = mv_{r}{\bf e_{r}} +
mv_{z}{\bf e_{z}}+m{\bf\omega}\times{\bf R }, \eqno(9) $$ where
$v_r=sin\theta (dR/dt)$ and $v_z=dz/dt$.

Differentiating this equation and taking into account $d{\bf
e_i}/dt = {\bf\omega\times e_i}$, $i = r, z $: $$ \frac{d{\bf
P}}{dt} = \left(\frac{d{\bf P}}{dt} \right)_ {n} -m\omega^2 r{\bf
e_r}+\omega\left(mv_r+\frac{d(mr)}{dt}\right) {\bf
e_{\varphi}},\eqno(10) $$ where $\left(\frac{d{\bf P}}{dt}
\right)_ {n}$ denotes the time derivative of the momentum defined
in the non-inertial frame of reference.

For the inertial forces (taking into consideration the
relationship: $r=Rsin\theta$) we have: $$ {\bf F_{in}} = m\omega^2
Rsin\theta{\bf e_r}-\Omega\left(mv+\frac{d(mR)}{dt}\right) {\bf
e_{\varphi}}.\eqno(11) $$

Generally the force responsible for BBW is the projection of ${\bf
F_{in}}$ on the direction perpendicular to the magnetic field
line: $$ F_{\perp} =
\left[F^2_{inr}cos^2\theta+F^2_{in\varphi}\right]^{1/2},
\eqno(12a) $$ where $$ F_{inr}=m_0\gamma\omega^2 rsin\theta,
\eqno(12b) $$ $$ F_{in\varphi}=-m_0\Omega\left(\gamma
v+\frac{d(\gamma R)}{dt}\right).\eqno(12c) $$ Note that (12a), in
the limit $\theta=90^0$, reduces to the expression (\cite{rm00}):
$$ F_{\perp} =
m_0\omega\left(2\gamma\frac{dR}{dt}+R\frac{d\gamma}{dt} \right).
\eqno(12d) $$ coinciding with the Coriolis force acting on the
particle.

While the electron moves along the magnetic field line, apart from
the inertial forces, it also experiences the Lorentz force: $$
{\bf F_L} =  \frac{q}{c}\bf{v_{rel}}\times \bf{B}, \eqno(13) $$
where $v_{rel}$ is the  velocity of the electron relative to the
magnetic field line and  $m_0$ and $q$ are the electron's rest
mass and charge, respectively. By virtue of the Lorentz force, the
electron gyrates around the magnetic field line and, during the
course of motion, the Lorentz force changes from parallel to
antiparallel to ${\bf F_{\perp}}$. If the Lorentz force is greater
than $F_{\perp}$ the particle moves along the magnetic field line.
The situation changes when $F_{\perp}$ exceeds $F_{L}$, in this
case ${\bf F_{\perp}}$ forces the particle to come off the
magnetic field line. Therefore, the assumption that the particle
sticks to the field line is no longer valid and the BBW becomes
important as a limiting mechanism for the centrifugal
acceleration.

From the  condition  $F_L \approx F_{\perp}$ (\cite{rm00}) one can
derive an estimate of the maximum Lorentz factor limited by the
BBW: $$ \gamma^{FB}_{max}\approx
A_1+[A_2+(A_2^2-A_1^6)^{1/2}]^{1/3}+[A_2-(A_2^2-A_1^6)^{1/2}]
^{1/3},\eqno(14a) $$ where $$
A_1=-\frac{ctg^2\theta}{12\widetilde{m}^{1/2}},\eqno(14b) $$ $$
A_2=\frac{q^2 l_e\times
L_{Edd}}{4m^2_0\widetilde{m}^{1/2}c^5}+A_1^3,\eqno(14c) $$ which
in the limit $\theta=90^\circ$ reduces to the form given in
(\cite{rm00}): $$ \gamma^{BBW}_{max}\approx \frac{1}{
\widetilde{m}^{1/6} }\left(\frac{B(r_L)q}{2m\omega
c}\right)^{2/3},\eqno(14d) $$ where $B(r)$ is the equipartition
magnetic field strength (which means that magnetic field and
radiation energy densities are equal) at the radius $r$ and is
given by the formula (\cite{rm00}): $$ B^2(r) = \frac{2l_e\times
L_{Edd}}{r^2c}.\eqno(14e) $$

%______________________________________________________________

\section{The Case of magnetic field lines fixed in the equatorial plane}

We start by considering the case when $\theta=90^\circ$. From
comparison of graphs of $\gamma^{BBW}_{max}(L)$ and
$\gamma^{ICS}_{max}(L)$ (see Fig.1) one can see that for AGN with
luminosities $L > 8\times 10^{40}erg/s$ the maximum Lorentz factor
attainable by electrons via ICS (solid line) is less than the
corresponding Lorentz factor attainable via BBW (dashed line),
which means that the latter is unimportant and it becomes
significant only for objects with $L<8\times 10^{40}erg/s$. From
Fig.1 it is also clear that, for $L>8\times 10^{40}erg/s$,
increasing the luminosity power, the maximum attainable Lorentz
factors decrease. From Eq. (8) we also can see that the maximum
Lorentz factors scale as $\gamma^{ICS}_{max}\sim 1/U_{rad}^2$, as
a result of the dependence of $t_{cool}$ on $U_{rad}$. The
conclusion made by Rieger and Mannheim (\cite{rm00}) is quite
different. These authors estimated the upper limit of the maximum
Lorentz factor attainable under the BBW for $l_e<10^{-3}$ (when
according to Rieger and Mannheim (2000) ICS is unimportant) as
$\gamma_{max}\sim1000$ and even for the highest possible magnetic
field strength - $B(r_L) = 100G$ - as $\gamma_{max}\sim2500$. The
correct value of the corresponding Lorentz factor is
$\gamma_{max}\sim4\times10^8$. This circumstance changes the
result of Rieger and Mannheim, because this value of
$\gamma_{max}$ is larger by many orders of magnitude, than the
corresponding limit of $\gamma_{max}$ given by ICS (equal, in this
case, to $\sim3.6\times 10^3$). This means that the latter is
dominant and the lower limit of $l_e$ at which the ICS still works
must be shifted from $10^{-3}$ to $8\times10^{-6}$ (see Fig.1).
The corresponding luminosity shifts from $10^{43}erg/s$ to
$8\times10^{40}erg/s$. In particular, if one considers the case $L
= 2\times10^{42}erg/s$ (when following Rieger and Mannheim (2000)
BBW must be important), using Eqs. (2), (6), (12d) and (13) one
can easily show that, during the course of motion, the Lorentz
force always exceeds Coriolis force and the maximum Lorentz factor
is limited only by  ICS (see Fig.2, Fig.3). Moreover, from Fig.2
it is clear that, at the beginning, the electron acceleration time
scale is shorter than the cooling time scale, which means that the
electron will accelerate, but, as soon as the Lorentz factor of
the particle becomes of the order of $10^5$, the electron energy
gain is counter-balanced by the losses via ICS and further
acceleration is impossible.

\begin{figure}
  \resizebox{\hsize}{!}{\includegraphics[angle=0]{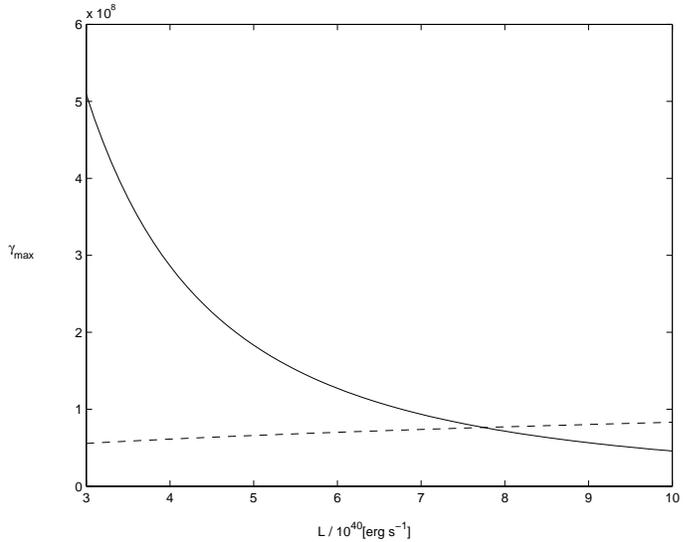}}
  \caption{Graphs of dependence of maximum Lorentz factors on $L$.
$\gamma^{ICS}_{max}$ and $\gamma^{BBW}_{max}$ are presented by the
solid line and the dashed line respectively. The set of parameters
is: $\tau = 1$, $\omega r_0/c = 0.4$, $v_0/c = 0.6$ and $\omega =
3\times10^{-5}s^{-1}$.}\label{fig1}
\end{figure}

\begin{figure}
  \resizebox{\hsize}{!}{\includegraphics[angle=0]{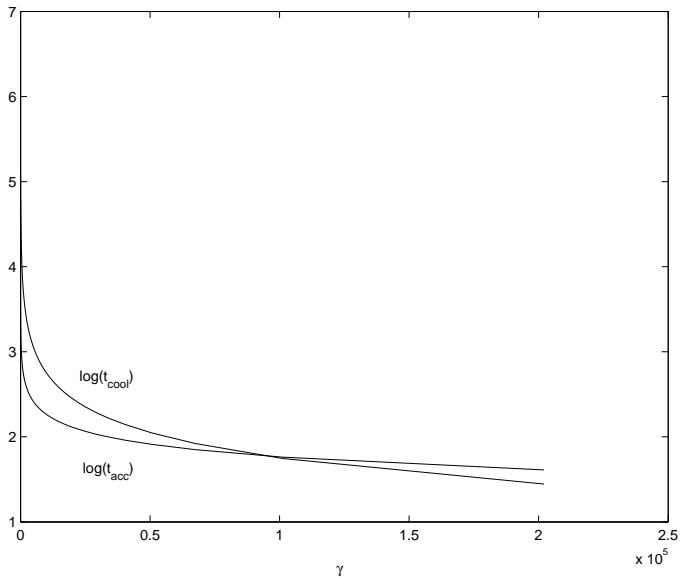}}
  \caption{Graphs of dependence of logarithms of time scales on
the Lorentz factor. The set of parameters is the same as in Fig.1,
except $l = 5\times10^{-4}$.}\label{fig2}
\end{figure}

\begin{figure}
  \resizebox{\hsize}{!}{\includegraphics[angle=0]{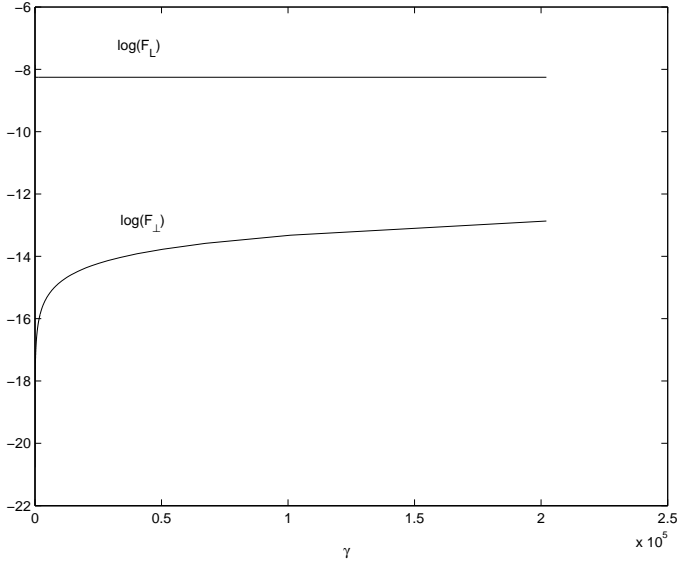}}
  \caption{Graphs of dependence of logarithms of the Lorents $F_{L}$
and Coriolis $F_{\perp}$ forces on Lorentz factors. The set of
parameters is the same as in Fig.2.}\label{fig3}
\end{figure}

\begin{figure}
  \resizebox{\hsize}{!}{\includegraphics[angle=0]{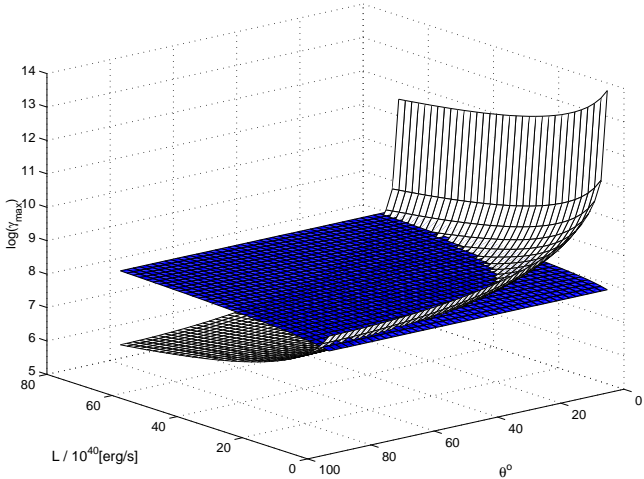}}
  \caption{Three dimensional graphs of $log\left(\gamma^{ICS}_{max}(\theta,l)\right)$
(bright surface) and
$log\left(\gamma^{BBW}_{max}(\theta,l)\right)$ (dark surface). The
set of parameters is the same as for Fig.1 except $L$ and $\theta$
which vary respectively in ranges:$[5\times 10^{40}, 7\times
10^{41}]egr/s$ and $\theta\in[0.1^0,90^0]$.}\label{fig4}
\end{figure}

Inverse Compton scattering of $0.1 keV$ energy photons produces
gamma rays with a maximum energy (\cite{rm00}) $$
\epsilon_{ph}\approx\frac{\gamma_{max}^2}{10^{10}}TeV. \eqno(15)
$$ Considering, as an example,  a luminosity power of $2\times
10^{42}erg/s$, we have seen that the maximum Lorentz factor is
$10^5$, which gives a maximum gamma ray energy of $\sim1TeV$.

Our approach is based on the assumption that electrons co-rotate
with the spinning AGN, which is valid only inside the Alfv{\'e}n
zone, because in this region the magnetic field is very strong
and, as a result, the flow follows it. On the other hand,
particles reach their upper energy limit in a region very close to
the light cylinder, which means that our approach is valid if the
following condition is satisfied: $(r_L-r_A)/r_L\ll 1$ (where
$r_A$ is the Alfv{\'e}n radius). $r_A$ can be estimated by the
expression: $B(r)^2/(8\pi)\approx m_0nc^2\gamma(r)$, where $m_0$
and $n$ are respectively the relativistic electron's rest mass and
the density. If one considers a range of $n_J/n_m = [10^{-4};
10^{4}]$, where $n_J$ is the jet bulk density and $n_m$ - the
medium density (suppose that $n_m\approx 1cm^{-3}$), by using Eqs.
(5) and (14e) one can show that for $L = 2\times 10^{42}erg/s $,
$(r_L-r_A)/r_L\sim [10^{-11};10^{-3}]$, which means that
co-rotation is valid almost for the whole course of motion.

% Usually jets are supposed to be under dense with
%respect to the medium: $n_J/n_m<1$, .

%______________________________________________________________

\section{The case of the inclined magnetic field lines}

In this section we examine how the situation changes when varying
the inclination of magnetic field lines.  For simplicity we
consider the case of straight magnetic field lines.

The critical luminosity power (the luminosity power at which both
mechanisms give the same maximum Lorentz factors) changes
depending on the inclination angle. From the graphs of
$log\left(\gamma^{ICS}_{max}\right)$ and
$log\left(\gamma^{BBW}_{max}\right)$ [see Eqs. (8) and (14a)] one
can see that for a luminosity less than $8\times 10^{40}erg/s$
(see Fig.1) the surface of $log\left(\gamma^{ICS}_{max}\right)$ is
over the corresponding surface of the BBW, independent of the
inclination angle (see Fig.4). Thus, with a luminosity of
$10^{41}erg/s$, the maximum Lorentz factor attainable by the
electron is limited only by the BBW. For higher luminosity AGNs
the BBW also may be important, but for relatively smaller angles
(see Fig.4).

For TeV blazars like Mrk421 and Mrk501 we have ultra-high energy
$\gamma$-ray emission in the range $[1-20]TeV$. The luminosity
power of Mrk421 and Mrk501 is estimated as $L\sim 10^{44}erg/s$
(\cite{bick}). If we consider magnetic field lines inclined by the
angle $\theta$, then using Eqs. (8) and (15) one may express the
mentioned angle in terms of $\epsilon_{ph}$:

\begin{figure}
  \resizebox{\hsize}{!}{\includegraphics[angle=0]{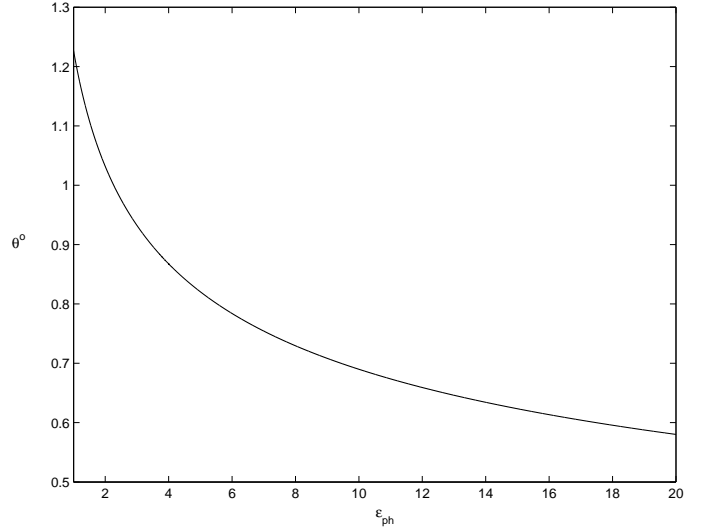}}
  \caption{Graph of dependence of $\theta(\epsilon_{ph})$ on $\epsilon_{ph}$.
  The set of parameters is: $\omega =
3\times10^{-5}s^{-1}$ and $L\sim 10^{44}erg/s$.}\label{fig5}
\end{figure}

$$ \theta^\circ \approx\frac{4\times 10^3c^3}{\omega
L}\times\left(\frac{\epsilon_{ph}}{10^{18}}\right)^{-1/4},\eqno(16)
$$

In Fig.5 we show the graph of $\theta(\epsilon_{ph})$. The set of
parameters is following: $\omega = 3\times10^{-5}s^{-1}$ and
$L\sim 10^{44}erg/s$. As it is seen from this figure, while
considering magnetic field line's inclination in the range
$\sim[0.6^\circ-1.2^\circ]$, one can explain the observed
ultra-high energies from Mrk421 and Mrk501. As one can also see,
by decreasing the inclination angle the corresponding energy of
the photon increases. To understand the behaviour shown in Fig.5,
one has to substitute the following expressions $R_L\approx
r_L/sin\theta$ and $\Omega = \omega\sin\theta$ into Eqs. (2) and
(6), then one gets $t_{cool}/t_{acc}\sim \Omega^{-1}$, which means
that for smaller angles the ratio of time scales is higher. As a
result, the particle will have more time for acceleration before
energy saturation and the energy gain will be higher. That is why
$\gamma_{max}^{ICS}\sim1/\Omega^2$ [see Eq.(8)], thus for smaller
angles the result of acceleration is more effective than for
higher inclinations, where acceleration is stronger but saturation
appears very soon, hampering subsequent energy gain for particles.
As a result of the angle dependence of
$\gamma_{max}^{ICS}(\theta)$, energies of produced photons become
higher for smaller inclinations [see Fig.5 and Eq. (16)].

The dependence on the angle is also very useful for studying
different spinning rates. In our analysis, we considered a certain
value of the angular velocity, but, on the basis of the above
consideration one can predict what may happen for other values of
$\omega$. As we have already seen, $\omega$ and $\theta$ are
equivalent, in the sense that dynamics of particles depends on
$\Omega$, but not on $\omega$ or $\theta$ independently. By
applying (8) and (15) one may see $\epsilon_{ph}
\sim(\omega\sin\theta)^{-4}$, which means that for lower spinning
rates the corresponding energies of produced photons are higher.
This happens because for lower angular velocities particles have
more time for acceleration and their resulting energy may be
higher (see the corresponding discussion about the inclination
angles).

%By applying the Eqs. (8, 15) one may see $\epsilon_{ph}
%\sim(\omega\sin\theta)^{-4}$, which means that for smaller
%spinning rates the corresponding energies of produced photons are
%higher. The reason of this is that, for smaller angular velocities
%the particle will have more time for acceleration and the
%resulting energy will be higher (see the corresponding discussion
%about the inclination angles).

%______________________________________________________________

\section{Conclusions}

In the present work we focused on the ultra-high emission from
AGN, the role of the centrifugal acceleration in this process and
the efficiency of different mechanisms limiting this acceleration.
We re-examined and generalized the problem considered in Rieger
and Mannheim (2000), where the authors argued that together with
the ICS, is also BBW an efficient limiting mechanism for electron
acceleration. They argued that the BBW limits $\gamma_{max}$ to a
few thousands, which is much less than $\sim10^5$ (the minimum
value of the Lorentz factor needed to produce TeV photons).

Revisiting the same problem  we showed that for AGNs considered by
Rieger and Mannheim (2000) as the acceleration limiting mechanism,
only ICS is important and consequently the upper limit of the
maximum Lorentz factor may be much more than several thousands. In
particular, we have shown that when the magnetic field lines are
straight and inclined to the axis of rotation by the angle
$\theta=90^\circ$, for AGN with luminosity $L>8\times
10^{40}erg/s$, ICS is dominant and BBW becomes significant only
for AGN with lower luminosities $L<8\times 10^{40}erg/s$.
Therefore, the example of an AGN with a luminosity power $2\times
10^{42}erg/s$, when as it was shown only ICS is important (for
$\theta=90^\circ$), we calculated the approximate value of the
maximum Lorentz factor of $10^{5}$ and argued that by virtue of
the ICS electrons with such Lorentz factors can produce $1TeV$
energy $\gamma$-rays.

We also considered the dependence of the validity of the mentioned
mechanisms on the inclination angle of the magnetic field lines
with respect to the rotation axis and have shown that also in this
case, for a wide range of possible inclination angles, ICS is the
most important mechanism limiting the Lorentz factor. The latter
value may reach up to $\sim10^8$. BBW becomes dominant mostly for
relatively lower luminosities with relatively higher angles, or
with higher luminosities and smaller angles. The exception is the
case when the luminosity is less than $8\times 10^{40}erg/s$, when
independently of the inclination angle the BBW is dominant. We can
estimate by Eq. (14a) that the minimum value of
$\gamma_{max}^{BBW}$ also is very high, $\sim10^7$.

Letting magnetic field lines have inclination angles in the range
$\sim[0.6^\circ - 1.2^\circ]$, we have shown that the
corresponding energies of photons due to ICS are of the order of
$[1-20]TeV$ which is sufficient to account for the emission of TeV
blazars.

For the wide range of AGN with two considered mechanisms, ICS
still remains the dominant factor in limiting the maximum Lorentz
factor of accelerated particles, letting them produce very high
energy photons $[1-20]TeV$. For low luminosity power AGN, at the
other hand, BBW may be significant but it also may provide high
maximum Lorentz factors of accelerated electrons.

An important restriction of our approach is that we studied only
straight magnetic field lines, whereas in realistic astrophysical
situations the magnetic field lines are curved. Therefore, it
would be interesting to generalize our approach to how the
considered problem changes, when the curvature of the magnetic
field lines is taken into account. It might be especially
important if we consider particle dynamics on a longer time scale,
over large length-scales, when the curvature of the field lines
cannot be neglected. A mathematical formalism for such a study
does exist (\cite{r03}) and it can be applied to the problem under
consideration.

The next limitation is related to the fact that the magnetic field
is not influenced by particle motion up to the region  in the
immediate vicinity of the light cylinder, where the magnetic field
is affected by the motion of the plasma. Our model has been based
on the study of the dynamics of a single particle, neglecting
complex processes in real relativistic plasmas. We must generalize
our approach and examine a more realistic model.

\begin{acknowledgements}

The research of Andria Rogava and Zaza Osmanov was supported in
part by the Georgian National Science Foundation grant
GNSF/ST06/4-096. Zaza Osmanov is grateful to Osservatorio
Astronomico di Torino for the hospitality and support. Andria
Rogava wishes to thank Katholieke Universiteit Leuven (Leuven,
Belgium) and Abdus Salam International Centre for Theoretical
Physics (Trieste, Italy) for supporting him, in part, through a
Senior Postdoctoral Fellowship and Senior Associate Membership
Award, respectively.

\end{acknowledgements}

\end{document}